## Digital Humanities in the TIME-US Project: Richness and Contribution of Interdisciplinary Methods for Labour History

*Marie Puren*

Marie Puren is an associate professor in history and digital humanities at the LRE at EPITA (Paris). She is an associate researcher at the Centre Jean-Mabillon (Ecole nationale des chartes).

In 2015, the *Annales* journal, traditionally open to interdisciplinary approaches in history, referred to 'the current historiographical moment [as] call[ing] for an experimentation of approaches'.[1] Although this observation did not exclusively refer to the new possibilities offered by the technological advancements of the time — particularly in the field of artificial intelligence[2] — it was nonetheless motivated by these rapid and numerous changes, which also affect the historiographical landscape. A year earlier, Stéphane Lamassé and Philippe Rygiel spoke of the 'new frontiers of the historian', frontiers opened a few years earlier by the realisation of the unprecedented impact of new technologies on historical practices, leading to a 'mutation des conditions de production et de diffusion des connaissances historiques, voire de la nature de celles-ci' ('transformation of the conditions of production and dissemination of historical knowledge, and even the nature of this knowledge').[3] It was in this fertile ground, conducive to the cross-fertilisation of approaches, that the TIME-US project was born in 2016. TIME-US is directly the result of this awareness and reflects the transformations induced by major technological advancements, disrupting not only our daily practices but also our historical practices.

---

[1] Annales 2015, 216.

[2] For example, convolutional neural networks, which have revolutionised the field of artificial intelligence, began to gain popularity just before the 2010s.

[3] Translated by the author. Lamassé and Rygiel 2014.

At the origin of the project, there was first a scientific objective: to reconstruct the remuneration, working hours, and wages of men and women in the textile industry in four major French industrial regions over the long term (from the late seventeenth to the early twentieth century). To achieve this goal, it was necessary to find ways to quantify these working hours and to better understand the remuneration definition process, which can be expressed in very different ways depending on the period, the remunerated task, and the source used. In other words, historians wanted to have access not only to data sets that they could study systematically and compare with data sets from (for example) other national contexts, but also to the possibility of exploiting qualitative sources using quantitative tools.

For France, the data available to researchers on women's time budgets and remuneration were particularly sparse.[4] One of the main challenges of TIME-US was to address this gap by creating data that would allow for a better understanding of women's work and their contribution to industrial development in the textile sector, and to compare them with their male counterparts. Gender historians have shown that this participation has been systematically under-recorded, and that it was necessary to use new sources to find tangible traces of this female activity.[5] The invisibility of women's work goes hand in hand with the idea that work has often been conceptualised as a remunerated task or one that was officially recognised as employment[6]. Indeed, women's work in the past has often (if not in most cases) been unpaid or not recognised as such (e.g., domestic tasks or 'help' provided by the wife in her husband's business), making it not easily identifiable by classic indicators such as quantified remuneration or official recording of professional activity.[7]

---

[4] Chagué, Le Fourner, Martini and Villemonte de la Clergerie 2022.
[5] See Humphries and Sarasúa 2012; Schmidt A. and E. van Neverdeen Meerkerk 2012; Ågren 2018a.
[6] See Sarti, Bellavitis and Martini 2018 on new estimates of women's work in censuses and the issue of unpaid work.
[7] Ågren 2018b, 226-7.

To quantify women's work in the past, labour historians cannot rely on the classic sources of their discipline, which allow to produce large statistical data series, systematically treatable in the form of databases. What to do when such data are not available? Should the task simply be abandoned? As Maria Ågren points out, the invisibility of women's participation in the labour market does not mean non-existence[8]; there must therefore be traces of it. To quantify women's economic activity, Sara Horrell and Jane Humphries, for example, turned to household budgets from 59 different sources (from Parliamentary Papers to autobiographical texts), which had never before been systematically used to identify women's work patterns and their contribution to family income.[9] In her study *A Bitter Living: Women, Markets, and Social Capital in Early Modern Germany* published in 2003, Sheilagh Ogilvie used information contained in court records to identify activities carried out by women and the time spent on these activities. Court records were not intended to record such information; yet, in their testimonies, witnesses often described in detail the activities they were engaged in while a crime was unfolding before their eyes. Sheilagh Ogilvie thus identified nearly 3000 such observations.[10]

These works have opened two main avenues for the TIME-US project. First, making already digitised sources accessible in homogeneous corpora.[11] Following the example of previous research, TIME-US mobilised varied sources containing traces of professional activities carried out by women in France during the period studied: these include both printed (posters and petitions, working-class newspapers, and contemporary surveys on workers) and handwritten sources (labour court decisions, police reports, company archives, personal archives, surveys, petitions).[12] One of the project's objectives was to gather and provide access to original data for France, taking advantage of the possibilities offered by the massive digitisation of heritage documents and the new tools offered by digital humanities. The creation of a processing chain for a heterogeneous digital corpus, with a view to its online publication, is thus one of the essential contributions of the project. As Sara Horrell and Jane Humphries noted, none of the sources used in their study, some of which were well known to historians, had 'been systematically analysed to reveal patterns in women's work and variation in the contribution of women [...] to family incomes across sectors and over time during industrialisation'.[13] This is also the case for the corpus assembled by TIME-US; but thanks to digital humanities, the project was able to provide access to the exploited corpus, offering a form of 'decentring' of the perspective on its sources and opening new avenues for exploration.

---

[8] Ågren 2018a, 144.
[9] Horrell and Humphries 1995.
[10] Ogilvie 2003.
[11] Translated by the author. Chagué, Le Fourner, Martini and Villemonte de la Clergerie 2022.
[12] Puren, Chagué, Martini, Villemonte de La Clergerie and Riondet 2018.

The second contribution of the TIME-US project lies in the exploitation of these digitised sources with tools and methods from digital humanities, and more generally from the field of computer science. One of the central challenges of the project was to obtain data analysable with quantitative methods, specific to labour, economic, or social history. But obtaining serial data from these sources is not trivial: firstly because the sources used do not always express this data in a directly quantified and quantifiable manner (working hours can be, for example, expressed circumstantially); secondly because the data bearing information for the historian are not easily identifiable by the human eye (how to systematically spot remunerations when they are, for example, expressed throughout a text?); finally, because this data can be 'scattered' in 'non-standard' sources for labour and economic history. When looking at work in the past, whether by women or men, one must be aware that relying solely on the possible 'quantified' traces available to the researcher is insufficient. These traces often take a linguistic form, which is therefore not directly quantifiable. While digitisation provides access to more sources, the historian is still faced with the same methodological problem: how to use quantitative methods to study these sources when they do not seem to lend themselves to this type of analysis?

---

[13] Horrell and Humphries 1995, 90-1.

In 1965, one of the pioneers of data quantification for the history of the bourgeoisie, Adeline Daumard, in her article 'Données économiques et histoire sociale', wrote:

> Certes tout ce qui relève de la description sociale n'est pas mesurable, mais un des objectifs de l'historien est d'étendre au maximum le champ de la statistique, même à des domaines qui paraissaient, avant ces tentatives, totalement irréductibles à une appréciation chiffrée.

> Certainly, not everything related to social description is measurable, but one of the historian's objectives is to extend the field of statistics as much as possible, even to areas that seemed, before these attempts, totally irreducible to a quantified appreciation.[14]

Several decades later and fully aware of the criticisms levelled at the wave of optimistic quantification in social history in the 1960s, this was also one of the objectives of TIME-US, which aimed to produce quantitative data from a vast textual corpus. In this, it is directly in line with quantitative history, which had its golden age in France in the 1960s and 1970s. But how to produce such data from numerous, disparate, textual, and poorly (if at all) structured sources? Drawing inspiration from the work of Sheilagh Ogilvie, Maria Ågren proposed using what she calls the 'verb-oriented method' to overcome this problem[15]. This method posits that, in fact, labour historians predominantly use textual sources, 'and that in texts it is the job of verb phrases to describe what people do'.[16] It therefore proposes to systematically identify sentence segments describing an activity carried out by people in the past. This kind of undertaking requires finding a method for systematically locating these segments in large sets of documents. Such tasks are particularly well-suited to computers; that is why TIME-US turned to natural language processing to handle large quantities of texts, annotate and structure this vast corpus of historical sources, and extract the necessary knowledge to achieve the TIME-US project's objective.

---

[14] Translated by the author. Daumard 196, 62.
[15] Ågren 2017.
[16] Ågren 2018b, 226.

It is this dual contribution of TIME-US that will structure this chapter. We will attempt to show how the fundamentally interdisciplinary approach of TIME-US, combining methods from history and computer science, has produced rich and new results, and opened up new avenues of research.[17] While it is a history project, TIME-US is also a digital humanities project, due to its interdisciplinary and experimental nature. It was also about 'inventing' new methods to analyse an unprecedented corpus of sources in light of the historical method. In a first part, we will see how TIME-US symbolised the turn taken by what some call 'digital history'. The project is indeed emblematic of the desire to take advantage of the new digital corpora made available on the web, using the tools and methods of digital humanities, while renewing the traditional methods of quantitative history. In a second part, we will focus on how TIME-US utilised the tools and methods at its disposal. We will see that TIME-US, despite the inherent constraints of such a project, particularly in terms of data extraction, adopted robust and proven computer tools while being aware of their contributions and limitations.

**Towards the 'Datafication' of Sources in Labour History**

<u>Digitised Historical Sources: A New Eldorado for Labour History?</u>

In 2023, *Gallica*, the digital library of the National Library of France, celebrated its twenty-fifth anniversary with the online release of its 10 millionth document.[18] This achievement is the result of extensive digitisation policies implemented by heritage institutions, spurred by the explosion of the Web in the latter half of the 1990s.[19] Historians now have an ever-growing number of digitised historical sources at their disposal, increasingly available in machine-readable formats.[20] Social history and labour history have also benefited from this massive wave of digitisation: by the late 2000s, the interest in digitising an increasingly diverse range of collections became apparent to conservation institutions dedicated to labour history research.[21] In 2014, the Social History Portal was launched by the International Association of Labour History Institutions (IALHI)[22], providing access to 900,000 digitised items, including archives, books, films, and posters.

---

[17] Chagué, Le Fourner, Martini and Villemonte de la Clergerie 2022.
[18] See the post '10 millions de documents et 25 ans d'existence' (30 mars 2023) : https://gallica.bnf.fr/blog/30032023/10-millions-de-documents-et-25-ans-dexistence?mode=desktop.
[19] Bardiot and Ruiz 2022.

Traditionally, labour historians rely on five main types of data: 'wages, occupational descriptors, work activities, material objects, and labour relations'.[23] It is rare for a single historical source to be able to provide all this data, which is why it is necessary to use a variety of sources. This is one of the revolutions brought about by what is called 'datafication', defined as 'the production of and the shift toward digital representations of historical sources as a prerequisite for storage, access, and analysis, not to mention their transmission and publication online'.[24] Digitisation indeed offers access to vast quantities of historical data, facilitating the cross-referencing of sources and the creation of diverse study corpora. TIME-US was thus able to utilise already digitised sources, collecting 8,000 image files from Internet Archive corresponding to thirteen volumes of the series of family monographs compiled in *Ouvriers des deux mondes* and *Ouvriers européens*[25], and 360 Lyon workers' newspapers from the nineteenth century downloaded from the *Numelyo* digital library. Notably, TIME-US itself contributed to this 'datafication' movement, with nearly 10,000 photographs taken during research conducted in the Lyon and Paris region archives.[26]

---

[20] Salmi 2021, 9.
[21] Van der Werf-Davelaar 2008.
[22] Blum 2014.
[23] Ågren 2020, 8.
[24] See the presentation of the conference 'Datafication in the Historical Humanities. Reconsidering Traditional Understandings of Sources and Data' (2-4 June 2022): https://datafication.hypotheses.org/
[25] For more information on these documents, see Hincker 2001.
[26] Chagué, Le Fourner, Martini and Villemonte de la Clergerie 2022.

The TIME-US project was initiated at a pivotal moment: major digitisation programmes had already borne fruit, providing access to document corpora previously inaccessible, and data that could be studied from a digital history perspective, that is, with 'computer-assisted ways'.[27] TIME-US is indeed in line with previous labour history projects relying on digitised sources. For example, in 2004, the *Écho de la Fabrique* project began, emblematic for its use of digitised serial sources.[28] This project contributed to the digitisation of a significant corpus of Lyon workers' newspapers published under the July Monarchy, with the online publication of this corpus encoded in XML-TEI.[29] A year before the birth of TIME-US, the *Accordi dei Garzoni* project also began, focusing on 32 registers held by the State Archives of Venice containing approximately 55,000 apprenticeship contracts declared by various profession guilds to the Giustizia Vecchia between 1575 and 1772.[30] Like TIME-US, this project aimed to extract structured historical information from these digitised documents, make it accessible in an open information system, and then proceed to its analysis.

TIME-US can be seen as the synthesis of the two trends highlighted by these projects: the extraction of data from a semi-massive historical corpus and the online publication of part of this corpus encoded in XML-TEI.[31] The originality of TIME-US lies in the variety of sources that constitute its study corpus. While *Écho de la Fabrique* and the *Accordi dei Garzoni* project worked on a single type of source, TIME-US has fully embraced the variety of digitised sources available and which labour history can now fully exploit. As with the *Écho de la Fabrique* project, nineteenth-century Lyon workers' newspapers have been an essential resource for TIME-US. The massive digitisation of ancient press initiated in the 2000s has made it one of the most used digitised sources by 'digital historians'.[32] Added to this are printed and handwritten economic, and legal archives produced at different times. The project thus contributes to the creation of digital archive corpora that can be (re)used and analysed by labour historians in light of their research questions.

<u>Datafy the Past</u>

The notion of 'datafication' was popularised in 2013 by Viktor Mayer-Schönberger and Kenneth Cukier in their book *Big Data: A Revolution That Will Transform How We Live, Work, and Think*. They explain that 'To datafy a phenomenon is to put it in a quantified format so it can be tabulated and analysed'.[33] While the datafication of the world is accelerated by digitisation, it is not new; accounting did not wait for computers to quantify goods, record financial transactions, and enter these figures into account books. But with digitisation, more and more historians are also practising datafication: because they consult digitised documents online, or because they themselves create digital duplicates of the archive documents they consult (by photographing them, for example). But this goes beyond simply reproducing historical sources in digital form. As Frédéric Clavert explains, datafication in the context of historical research must primarily be seen as a 'process' that goes from digitising a document to analysing it with computer tools.[34] It is this process that leads from sources to data.

In historical research, the source always remains at the heart of historians' work, and it is rare that they consider their objects of study to be data. However, history is not unfamiliar with this notion of 'data'. We recall here the words of Adeline Daumard, who explained that 'one of the historian's objectives is to extend the field of statistics as much as possible'.[35] As early as the 1950s[36], but especially in the 1960s and 1970s[37], social, economic, and demographic history in France became aware of the possibilities offered by computers to analyse large quantities of data. Emmanuel Le Roy Ladurie and Pierre Couperie even saw it as a true revolution: speaking of Parisian rents from the late Middle Ages to the eighteenth century, they wrote in 1970 in the *Annales* that if until then

> l'énormité de la documentation semble avoir paralysé les chercheurs [...] Il semble, cependant, que le moment soit venu pour une nouvelle approche du problème : les techniques modernes, issues des ordinateurs, permettent une véritable révolution historiographique ; elles autorisent le traitement exhaustif d'un très grand nombre de données [...].

> the enormity of the documentation seems to have paralysed researchers [...] It seems, however, that the time has come for a new approach to the problem: modern techniques, derived from computers, are enabling a veritable historiographical revolution; they allow a very large amount of data to be processed exhaustively [...].[38]

[33] Mayer-Schönberger and Cukier 2013, 72.
[34] Clavert 2016, 120.
[35] Daumard 1965, 62.
[36] Clavert and Noiret 2013, 18.
[37] Salmi 2021, 9.

Using large amounts of information is not new in history - let us remember the works of Fernand Braudel in *La Méditerranée et le monde méditerranéen à l'époque de Philippe II* (1949) and *Civilisation matérielle, économie et capitalisme* (1977), which aimed to manipulate a vast reservoir of knowledge -, but the massive digitisation of historical sources changes the game. These new deposits of digitised archives constitute the starting point, the 'precondition' of so-called 'digital' history. It is indeed only with such material available in abundance that we can truly take advantage of new tools intended to study these large volumes of historical data.[39] Some, like Jo Guldi and David Armitage in *The History Manifesto* (2014), see it as a call to resume *longue durée* investigations that would have been abandoned in favour of research focused on micro-history. The deliberately provocative stance, taken by the authors of this 'manifesto', gave rise to particularly rich debate in the historical community, with *Annales* devoting an issue to 'Debating the *Longue Durée*'[40]. In 2016[41], Guillaume Calafat and Eric Monnet wondered whether we were not witnessing 'the return of economic history', notably with a comeback of works focusing on the *longue durée* - ranging from a few centuries to a millennium - which then constitutes the privileged time scale for understanding economic changes and transformations[42]. Without drawing specific conclusions, Maria Ågren in 2020 also drew a parallel between the revitalisation of interest in 'labour relations across time and space' and the growing interest in new technologies enabling the analysis of this large quantity of historical data, 'often discussed as "big data" or "digital humanities"'.[43]

---

[38] Translated by the author. Le Roy Ladurie and Couperie 1970, 1002-3.
[39] Salmi 2021, 9.
[40] See *Annales* 2015.
[41] The French version of the article was published in 2016.

A project of labour history but also of digital history, TIME-US seems to us to be the heir of a double intellectual lineage: that of quantitative history first embodied by the French Annales school and, on the other side of the Atlantic, by American cliometrics focused on quantitative analyses[44]; and that of humanities computing today identified under the term 'digital humanities'.[45] TIME-US emerges from this encounter between historians' desire to exploit these billions of digitised texts and the access to increasingly powerful analytical tools allowing the exploration of these historical big data.

<u>The Essential Contribution of Natural Language Processing in the 'Datafication' of the TIME-US Corpus</u>

While digitised and born-digital data were once rare due to the difficulty of production and sharing, researchers today face an abundance of data, directly accessible via a computer screen. However, this apparent availability masks the fact that these data are rarely, if ever, usable as they are.[46] Data can be 'messy' and therefore need to be cleaned; they can also be partial and biased; in some cases — such as digitised historical documents — data extraction is necessary to be able to process them. According to Christof Schöch's definition, data in the humanities are constructed through careful choices made by researchers; they are also 'abstractions' in a machine-readable and usable format that represent certain aspects of a given research object.[47]

---

[42] Calafat and Monnet 2017.
[43] Ågren 2020, 6.
[44] Blaney, Winters, Milligan and Steer 2021, 8.
[45] Crymble 2021, 45.
[46] Poibeau 2014.

Information extraction from an image is a prerequisite for most digital history projects. Many digitisation projects have primarily been designed to make frequently consulted documents more easily accessible, without immediately considering the possibility of analysing their content using a computer. This can create a certain frustration among social science researchers.[48] In the case of the TIME-US project, this was one of the main challenges faced by project members early on: how to extract the necessary information to answer our initial question when we have at our disposal corpora of texts in image format?

The contribution of computer science, and more specifically Natural Language Processing (NLP), is essential here. As Michael Piotrowski points out, many projects, like TIME-US, rely on digitised texts.[49] The use of information technologies has thus primarily focused on text: for example, the Text Encoding Initiative (TEI), one of the founding projects of digital humanities, has since the 1980s[50] provided recommendations for the creation and management in digital form of all types of data (initially textual) created and used by humanities researchers. More broadly, text analysis is the dominant trend in digital humanities, and thus in digital history.[51] This is because a vast majority of humanities disciplines use textual data. For digital history, it is also because there is a long tradition of historical linguistics that studies digitised corpora, long before they became sources in their own right for historians.[52] NLP offers methods and tools to exploit these large quantities of text, particularly in two areas: digitisation — and with it the extraction of texts from digitised images — and the processing of these texts with a computer. Michael Piotrowski goes even further by stating that if the humanities want to achieve solid results through the analysis of large quantities of texts and the use of quantitative methods, 'they will need NLP as a basis for all higher-level analyses.[53] In the context of digital history, he even makes NLP 'an auxiliary science of history, similar to archaeology, diplomatics, palaeography, etc., which are indispensable for evaluating and using historical sources'.[54]

---

[47] Schöch 2013.
[48] Poibeau 2014.
[49] Piotrowski 2012, 8
[50] Clavert and Noiret 2013, 19.
[51] Wevers and Smits 2020, 194.
[52] Salmi 2021, 43.

This obviously requires skills that historians do not traditionally possess. This situation calls for collaborations with computational linguists[55], as the TIME-US project did by bringing together a team of historians and specialists in digital humanities and NLP. Such collaboration offers benefits in both directions: it allows answering a historical question using large corpora of texts; it also advances the state of the art in NLP by providing 'specific use cases' and 'real-world problems'. NLP can then experiment and adapt its techniques to 'different languages than English, to different corpora than newspapers and to different periods than just the twenty-first century'.[56] This type of collaboration is indeed particularly fruitful for computational linguists because it offers them 'new, original and complex challenges'.[57] The collaboration between historians and computer scientists thus constituted a particularly interesting experience for the participants in the TIME-US project, as the NLP experts first had to understand the historians' expectations, and then design appropriate tools. This meant setting up a process of continuous discussion and exchange, while respecting a 'trial and error' method that allowed a degree of agility in setting up the processing chain[58]. For TIME-US, these efforts enabled the establishment of a genuine 'datafication' process, from text extraction to its online publication, through its annotation and structuring. The aim is to move from sources to data; in other words, producing machine-readable and human-usable data from a digitised historical document.

---

[53] Piotrowski 2012, 8

[54] Piotrowski 2012, 8

[55] Kemman 2021, 37.

[56] McGillivray, Poibeau and Ruiz Fabo 2020.

[57] McGillivray, Poibeau and Ruiz Fabo 2020.

[58] TIME-US was a test project at Inria (of which the author was a member from 2016 to 2018) for the deployment of a processing chain for digitised ancient documents. TIME-US particularly benefited from the work carried out by Thibault Clérice, to whom we extend our warmest thanks.

**For a Digital History of Labour**

<u>From Sources to Data: 'Datafying' the TIME-US Corpus</u>

The 'datafication' process implemented by TIME-US comprised four main steps: collecting digitized sources (constituting the study corpus), acquiring the text (which includes segmentation, transcription of images and text normalization), structuring[59], and finally online publication.

*Acquire*

The creation of the corpus is then followed by the acquisition of data from digital copies. This is an essential stage in analysing texts using NLP tools: not only for any quantitative analysis, but also for the online publication of a searchable corpus. In recent years, considerable efforts have been made to develop optical character recognition (OCR) systems for printed texts, and handwriting recognition (HTR) systems, both based on artificial intelligence. The ongoing development of these systems means that we can achieve perfectly satisfactory (but not error-free) performance on modern printed texts; however, the performance is much lower on older texts because there is little data available to train AI to recognize ancient characters.[60] The problem is even more complicated for handwritten texts, as handwriting recognition also requires training data, and thus manually transcribing a significant portion of the study corpus. In the field of digital history, the data studied are rarely the result of simple collection; data acquisition is a long process that requires significant effort. Johanna Drucker, speaking of humanities data, prefers to use the term 'capta', emphasizing the fact that 'Capta is "taken" actively while data is assumed to be a "given" able to be recorded and observed'.[61] Although text acquisition from digitized documents is not strictly one of the research areas of NLP[62], it is a crucial step for NLP too. Michael Piotrowski indeed points out that if digitization is of interest to NLP, it is not only because it provides the 'raw' data necessary for analysis. First, the quality of document digitization has a significant impact on the results of processing: if the quality is poor, text extraction will not be error-free, which will result in biased outcomes. Second, NLP has a role to play during the text acquisition process and their post-correction.[63]

---

[59] Chagué, Le Fourner, Martini and Villemonte de la Clergerie 2022.
[60] Blouin, Favre and Auguste 2022, 79.

The task is even more difficult when it comes to acquiring text from a non-homogeneous corpus. The TIME-US corpus indeed presents several typical challenges faced by digital humanities researchers. It consists of handwritten documents in non-contemporary French with technical vocabulary (in this case, from the textile industry). If the volume of texts to be studied is substantial enough to be difficult to exploit 'manually,' it is not strictly speaking a 'big data' corpus, but rather a medium-sized one whose analysis consisted of a more fine-grained exploration (rather than frequency-based). This corpus mainly comprises minutes of labour courts (5458 sentences), worker press (14,204 sentences), and monographs of *Ouvriers des deux mondes* and *Ouvriers européens* (113,933 sentences). Other documents (173,031 sentences) were also collected (police reports, commercial court, silk trade, dictionary, etc.). In the limited time available to TIME-US, it was impossible to annotate and publish the entire corpus. A 'double corpus' approach was then adopted: on the one hand, a homogeneous consultation corpus (consisting of two collections, newspapers from Lyon about labour courts audiences (641 news) and minutes from Parisian labour courts audiences (139 cases)) was processed and put online for detailed exploration by researchers; on the other hand, a broader and more diverse acquisition corpus served as a basis for acquiring domain-specific knowledge, including terminology and ontology.[64]

---

[61] Drucker 2011.
[62] This is one of the areas of computer vision and document processing.
[63] Piotrowski 2012, 25-7.

The text contained in image files was extracted using segmentation and automatic transcription tools: first with the Transkribus platform[65], then with Kraken[66] deployed online on the eScriptorium platform[67], both of which allow automatic recognition of document structure and automatic transcription of texts.[68] However, the process is not as simple as it seems. First, the system must be trained to recognize the page structure. This 'segmentation' step is absolutely necessary for the machine to recognize lines, then words, and finally characters. Users must train the model by indicating and then correcting the page segmentation. Only then will the model be able to recognize characters and extract text from the page. This process is not error-free: a certain number of pages often need to be corrected to achieve better results. An attempt at post-correction and normalization was also initiated. For example, *pardevant* was corrected to *par-devant* (before, in presence of), or *engagemens* to *engagements* (commitments).[69] A total of 5570 occurrences were corrected. Abbreviations have also been developed, and dates standardised using a system of rules and regular expressions.[70]

*Structure*

Text extraction is followed by a structuring phase of this data into XML-TEI files. The choice of XML-TEI format was guided by its importance in the digital humanities community. It has gradually become the preferred format for sharing humanities data, especially textual data.[71] XML is also a format that is widely used in NLP, as it makes it easy to structure the texts analysed with a layer of annotations. For example, it can identify 'tokens'[72], sentences or named entities (persons, places, organizations, etc.).[73] Thanks to the use of XML-TEI, TIME-US has been able to create a semantic annotation layer, enabling it not only to record the descriptive elements of documents (their metadata), but also to annotate certain parts of the text, such as concepts or named entities, which can then be used to explore the corpus. Unfortunately, the structures of the sources can vary considerably; it was nevertheless necessary to identify similar information and, therefore, to identify the same types of annotations. The objective was to design a common annotation model, which still takes into account the diversity of forms taken by the information. This annotation model was formalized using a TEI schema, 'qui permet de moduler la spécificité des structures de chaque ensemble et l'unité de l'annotation sémantique' ('which allows modulating the specificity of the structures of each dataset and the unity of the semantic annotation').[74] It is important to note that this schema was designed using a bottom-up approach. A subset of documents was first defined before being gradually expanded. During this phase, text portions were manually annotated to validate and modify modelling choices. Documents are annotated in XML-TEI using an automatic structuring workflow that respects the internal logic of the documents. Named entities are also automatically recognised.[75]

In a second phase, these XML-TEI documents are enriched using an annotation pipeline initially designed to process Agence France Presse (AFP) news.[76] This process begins with tokenization, including the detection of named entities, followed by alignment using standoff positions anchored at the token level. Parsing then facilitates the identification of multi-word concepts, which are subsequently enriched through entity extraction and linking, leveraging contextual reinforcement from nearby concepts. This method not only ensures the precise annotation of individual elements but also supports the comprehensive semantic structuring of the corpus. The workflow's efficiency is underscored by the significant scale of annotated data, which includes 780 documents, 17,510 sentences, and 413,186 tokens, leading to the detection of 28,503 entities and 40,200 concepts.[77]

---

[72] Tokens refer to smaller parts than the sentence: they can be words or even series of characters. It all depends on the analyses carried out.

[73] Piotrowski 2012, 60.

[74] Chagué, Le Fourner, Martini and Villemonte de la Clergerie 2022.

[75] Dupont 2017.

[76] Clergerie and Martin 2021.

*Publish*

The structured data in XML-TEI was finally published online, enabling historians - and anyone else interested - to explore this corpus in quite a detailed manner. The goal was to create an adapted consultation interface, exploiting rich semantic annotations. Barbara McGillivray, Thierry Poibeau and Pablo Ruiz Fabo emphasize that one of NLP's objectives is also to give access to its tools to domain experts from the humanities and social sciences, through dedicated interfaces.[78] This is why the TIME-US corpus consultation interface[79] was developed, enabling its users to explore the corpus by navigating between documents or formulating queries based on annotations. This interface exploits the modelling of the XML-TEI document, and the semantic annotations obtained through NLP, using TEI Publisher[80], a tool designed for publishing digital publishing projects.[81]

<u>Multiplying Reading Scales: Distant Reading / Close Reading / Blended Reading</u>

The TIME-US project seems emblematic of a dual trend: the necessary exploration of large corpora using quantitative methods - which now include NLP techniques - and the desire to conduct precise analyses of 'micro' phenomena. The corpus consultation interface allows both document-by-document navigation, and access to all occurrences of, for example, a particular concept, enabling exploration through either a qualitative approach (e.g., detailed analysis of the source of a concept) or a quantitative one (e.g., evaluating the frequency of a term's appearance). Alix Chagué, Victoria Le Fourner, Manuela Martini, and Eric Villemonte de la Clergerie explain that TIME-US indeed combined two complementary approaches:

> une approche micro-qualitative impliquant une analyse empirique très approfondie des contextes historiques de production des sources utilisées [et] une approche quantitative ancrée dans le champ des humanités numériques et mettant en relation historiens et historiennes et spécialistes des outils et méthodes informatiques.

> a micro-qualitative approach involving very thorough empirical analysis of the historical contexts of the sources used [and] a quantitative approach rooted in the field of digital humanities, connecting historians with computer tool and method specialists.[82]

---

[77] Clergerie and Martin 2021.
[78] McGillivray, Poibeau and Ruiz Fabo 2020.
[79] https://timeusage.paris.inria.fr/exist/apps/timeus-corpus/index.html
[80] http://teipublisher.com
[81] Clergerie and Martin 2021.

As Hannu Salmi explains, 'If the nature, quality, and extent of the source material available for research have changed, it is natural that the research toolbox must change in tandem'.[83] Access to large quantities of sources requires historians to adapt their working methods to explore this data. How is it possible, within the finite framework of a project like TIME-US, to fully exploit 360 worker newspapers and thousands of files from the *Ouvriers des deux mondes*? This is where the notion of 'distant reading' comes in, first appearing in 2000 in Franco Moretti's article 'Conjectures on World Literature'.[84] It involves abandoning what he calls 'close reading' to make 'a little pact with the devil: we know how to read texts, now let's learn how not to read them'.[85] He adds that 'the more ambitious the project, the greater must the distance be'.[86] A literary historian, Franco Moretti is particularly interested in literary analysis, highlighting the idea that this approach allows the identification of 'patterns' within literary corpora spanning large historical periods and vast geographical areas. Distance is 'a condition of knowledge: it allows you to focus on units that are much smaller or much larger than the text: devices, themes, tropes—or genres and systems'.[87]

---

[82] Chagué, Le Fourner, Martini and Villemonte de la Clergerie 2022.
[83] Salmi 2021, 33.
[84] Moretti 2000.
[85] Moretti 2000.
[86] Moretti 2000.

Digital humanities quickly embraced distant reading to apply it to other disciplinary fields, especially history. This approach seems particularly relevant when researchers are faced with massive document corpora and have computer tools to analyse this 'big data'. While Moretti did not specifically mention computing when he began developing the concept of distant reading, the popularization of this approach has promoted the return of statistical analyses and quantitative methods to explore digitized corpora.[88] If digital history has easily adopted distant reading, it is also because history had already 'read at a distance' historical sources several decades before Franco Moretti's works.[89] A French school of discourse analysis[90] emerged in the 1970s, where the research of Régine Robin or Jacques Guilhaumou flourished.[91] Antoine Prost's work on the *Vocabulaire des proclamations électorales de 1881, 1885 et 1889* in 1974 also marked several generations of historians and inspired today's digital historians. Firstly, because linguistic methods were used to address a historical issue.[92] Magali Guaresi explains that these studies initially aimed to 'observer différemment les corpus de façon à nourrir, baliser et encadrer l'interprétation du sens historique des archives textuelles' ('observe corpora differently to nourish, mark out, and frame the interpretation of the historical meaning of textual archives').[93] Secondly, because this work, which required significant efforts to make the compiled data readable by a computer[94], was carried out in collaboration with statisticians to apply then-new methods to analyse these texts.[95]

---

[87] Moretti 2000.
[88] Salmi 2021, 34.
[89] Lemercier 2015, 276.
[90] Dumont, Julien and Lamassé 2023.
[91] Guilhaumou 2010, 720-1
[92] Bardiot and Ruiz 2022.
[93] Guaresi 2019.

This study, which was particularly innovative at the time, both in terms of method and the question it posed, paved the way for experiments conducted by mixed teams of historians and computer scientists on large corpora of historical texts. TIME-US has also favoured such approaches, utilising distant reading methods, while also adopting a 'micro-qualitative' approach. While it is essential to take advantage of the possibilities offered by the proliferation of data and digital corpora, it is also crucial not to abandon the practice of 'close reading'. If distance is a condition of knowledge, it is because it allows a better understanding of the contexts of production of texts, and thus 'consider how the bigger picture might change our view of the details'.[96] It is therefore essential to go back and forth between distant and close reading, as many digital historians have emphasized.[97]

In our view, the condition of knowledge lies in this constant motion between these two scales. While the *longue durée* as advocated by Jo Guldi and David Armitage offers new research perspectives drawing on large digitized corpora, 'it did not seek to oppose other approaches', and such an opposition seems far 'excessively simplistic' as the *Annales* journal rightly pointed out, shortly after the publication of *The History Manifesto*.[98] In *Exploring Big Historical Data. The Historian's Macroscope* (2022), Shawn Graham, Ian Milligan, Scott Weingart, and Kim Martin prefer to propose the notion of 'macroscope' to show that it is not about opposing micro- and macro-history. Broadly speaking, one can say that microhistory involves working in depth on 'a single story or moment in history'[99], while macrohistory focuses on major trends and their development over the long term. The historian's 'macroscope' is not confined to macrohistory; it can indeed be used very pertinently in a microhistorical approach. The authors give the example of studying thousands of tweets posted online during a debate for the US presidential election: if a 'macroscope' is needed to study this data, the goal is to study in detail a single historical event.[100]

---

[94] It should not be forgotten that, at the time, for a text to be read by a computer, it had to be recorded by hand on a punched card. Each character was coded using a set of perforations.
[95] Bardiot and Ruiz 2022.
[96] Salmi 2021, 37.
[97] Salmi 2021, 37
[98] Annales 2015, 216.
[99] Graham, Milligan, Weingart and Marti 2022, 2.

One successful method would even be to play with more than two reading scales. The *Annales* thus emphasize that 'there exist a whole range of intermediary possibilities between the macro and the micro approaches, which Braudel in fact recommended exploring in order to recognize the complexity of histories and their temporal inscription'.[101] Karine Karila-Cohen, Claire Lemercier, Isabelle Rosé, and Claire Zalc also adopt the same view by advocating a 'diversité [...] des focales utilisées' ('diversity [...] of the focal lengths used').[102] They emphasize the need

> [de] combin[er] des points de vue, les allers et retours entre des cas singuliers et une structure globale dont on ne peut percevoir qu'une facette à la fois, [qui] permettent de multiplier les pistes interprétatives pour affronter les silences et l'hétérogénéité des sources.

> [to] combine viewpoints, back and forth between singular cases and a global structure that can only be perceived in one facet at a time, [which] allow multiplying interpretative paths to face the silences and heterogeneity of sources.[103]

According to the authors, this is also a way to reconcile 'différentes manières de faire de l'histoire' ('different ways of doing history').[104] Tim Hitchcock also suggests working on small

---

and large scales and considering everything in between[105], a method Alexander Stulpe and Matthias Lemke have dubbed 'blended reading'.[106]

We believe that TIME-US is a good example of a 'blended reading' project, combining both distant reading using NLP techniques, and close reading, particularly through the attention paid to the selection of sources. It is recalled that the project conducted 'une analyse empirique très approfondie des contextes historiques de production des sources utilisées' ('a very thorough empirical analysis of the historical contexts of the sources used').[107] A long and meticulous selection of sources has made it possible to compile

> un ensemble de sources pour une période longue sur les salaires et les revenus en nature des travailleurs et travailleuses, selon le temps d'exécution, les tâches accomplies, le type de rémunération, les périodes d'activité, le statut, le sexe, l'âge (dans la mesure du possible) en les situant dans leurs lieux de production (atelier, usine, domicile).

> a set of sources for a long period on the wages and income in kind of male and female workers, according to the time worked, the tasks performed, the type of remuneration, the periods of activity, the status, the sex, the age (as far as possible) by locating them in their places of production (workshop, factory, home).[108]

In this respect, TIME-US aligns with the perspective developed by Claire Lemercier, who explains that distant reading is by no means a 'quick and easy reading, performed by tools that would make the *longue durée* immediately accessible to historians'.[109] To ensure solid and relevant results, particular care must be taken in constituting the corpus, which must be built rationally, depending on the research question. It is therefore not about 'amalgamating the largest possible number of texts' but rather restricting the corpus, understanding its structure, knowing the different documents it comprises, and being aware of its potential biases.[110] In other words, it's about being a historian: the critical analysis of sources and their reasoned selection is undoubtedly one of the essential contributions of the historical method to distant reading, digital history and more generally digital humanities. Jo Guldi even calls on data scientists to see data as historians do: incomplete, full of bias, prejudice and even lies.[111]

---

[105] Hitchcock 2014.
[106] Stulpe and Lemke 2016.
[107] Translated by the author. Chagué, Le Fourner, Martini and Villemonte de la Clergerie 2022.
[108] Translated by the author. Le Fourner, Chagué, Martini and Albert 2022.
[109] Lemercier 2015, 277.

<u>In the footsteps of the 'verb-oriented method'</u>

The use of blended reading by TIME-US is based on a strong methodological foundation: the reality of past work can only be known through meticulous reading of texts that have recorded and described work-related activities. We have indeed seen that the recording of these activities could be done in very diverse sources, whose primary purpose was not to describe labour activities. The TIME-US corpus thus comprises manuals and surveys, individual memoirs and worker press, petitions and police reports, labour court records, bankruptcies, and contravention registers of guilds from the modern era, originating from the Paris, Lyon, and Lille regions.[112]

An inspiration for dealing with this corpus and, ultimately, answering the initial research question, is notably found in the work of Maria Ågren, who, (as said above) drawing inspiration from Sheilagh Ogilvie's work, developed the 'verb-oriented method' The historian of labour assumes that 'verb phrases' are 'concrete descriptions of actual work tasks' and describe what people actually do in their daily lives. If one no longer relies solely on classic indicators (remunerations and occupational titles) to identify these activities, it is then possible to better capture all the modalities taken by work in the past, regardless of the value or name given to these activities at a given time.[113] Moreover, verb phrases 'are attractive to

---

the historian because they are concrete descriptions of actual work tasks'[114]: in other words, it is the actors themselves or direct witnesses of these activities who describe, in their own words, what they themselves or others actually do.

The 'verb-oriented method' paved the way for TIME-US, which was able to adopt its methodological stance: it involves systematically identifying, in the sources, the parts of the text that describe work-related activities. To answer the research question posed by TIME-US, it is not sufficient to merely identify verb phrases. A first step therefore consisted of defining the pieces of information necessary to address the initial issue. It was then decided to extract three categories of information: information related to persons, entities; segments related to work and remuneration; entities and segments related to the expression of time.[115] It was also necessary to build a corpus that would provide enough information of this type to create relevant datasets: the constitution of the corpus is therefore essential, as we want to extract enough data to ensure their representativeness; we also want these data to be relevant to answer the research question, and of the same nature to make comparisons between the industrial regions studied.

Adopting an approach such as the 'verb-oriented method' requires setting up a process to identify and extract the verb phrases that interest the labour historian. It can indeed be assumed that, in a given text, not all verb phrases concern only work activities. If, for a human being, it may be easy to identify the expressions of interest in a single text, or even a small number of texts, it becomes much more difficult and even impossible when tackling large corpora of serial sources (such as the press). Omissions can quickly multiply, and errors can easily slip into the records. But as Maria Ågren points out, it is precisely when applying this method to 'huge datasets' that 'it becomes particularly strong; while the single

---

observation can be fragmentary and hard to understand and classify, large amounts of observations allow us to discern general patterns'.[116]

TIME-US decided to delegate the task of identifying relevant text segments to a computer. This is a form of 'distant reading' of the corpus, as the computer systematically identifies text segments that correspond to the sought pieces of information. More precisely, it performs what is called knowledge extraction tasks. This is a key research area in NLP, which has developed various techniques well-suited for exploring large document corpora. For TIME-US, FRMG[117], a wide-coverage grammar for French, was used to parse and annotate the corpus. The aim was to build semantic networks and extract semantic relations through patterns and vocabulary, enhancing the robustness of linguistic data. Terminology extraction focused on identifying over-represented nominal sequences with strong internal cohesion, following specific construction patterns. Examples include phrases like *chef d'atelier* (workshop manager) and *paires de bas* (pairs of stockings), which are analysed for their grammatical structure.[118] The distributional hypothesis, as proposed by Zellig S. Harris[119], suggests that semantically close words occur in similar syntactic patterns, and this principle was applied to group words into concepts within the corpus. Semantic Role Labeling (SRL) was employed to find roles and fillers associated with verbal predicates, and an unsupervised approach inspired by Open Information Extraction was used to extend syntactic paths and fillers.[120] This involved iterative steps of defining seed words, finding syntactic paths, and expanding seed words to refine the concept lists and annotate semantic relations effectively. These semantic networks allow, for example, the representation of links between a product and materials.[121]

---

[116] Ågren 2018b, 227.
[117] Villemonte de la Clergerie, Sagot, Nicolas and Guénot. 2009; Morardo and Villemonte de la Clergerie 2014.
[118] Clergerie and Martin 2021.
[119] Harris 1954.
[120] Clergerie and Martin 2021.
[121] Chagué, Le Fourner, Martini and Villemonte de la Clergerie 2022.

A detailed analysis revealed insightful distributions and representations across various categories. For instance, the entity distribution highlighted a predominance of person-related entities (14,797 occurrences) followed by locations (6085) and dates (2470). Similarly, concept distribution showcased 'agent' as the most frequent type with 12,916 occurrences, followed by concepts relating to 'products' and 'money'. The most representative concepts within these categories include roles such as *chef d'atelier* (workshop manager) and *ouvrier* (worker), job titles like *négociant* (trader) and *fabricant* (manufacturer), products such as *pièce* (part) and *façon*[122], and terms related to financial transactions such as *indemnité* (indemnity) or *somme* (sum). Additionally, the gender distribution analysis pointed to a clear under-representation of feminine entities, with significantly lower counts in all categories compared to their masculine counterparts.[123]

This automatic knowledge extraction does not undermine the use of close reading for TIME-US at all. An example developed by Maria Ågren particularly illuminates the approach adopted by the project. She focused on the expression 'to close the door.' There is a difference between 'to close the door because it is cold' and 'to close the door of the chicken coop.' While the second example likely corresponds to a task performed by a servant for their master, it is more difficult to determine whether the first example pertains to a work activity, or an activity related to well-being and/or self-preservation. It is therefore necessary to look at the context in which this expression was used and return to the source from which it originates. This example is not used to question the use of distant reading; on the contrary, as Maria Ågren emphasizes, 'Saving us time and money by swiftly identifying the majority of verbs of interest, language technology will allow the historians to spend more time on contemplating the odd and intriguing examples'.[124] It is with this perspective that the corpus consultation interface was implemented, enabling easy interplay between the two reading scales, moving from the corpus to the individual source, and vice versa.

---

[122] In the context of the corpus, this term refers to the work carried out by the person shaping an object, or to the fact of working for someone else without supplying the raw material ('*Travailler à façon*').
[123] Clergerie and Martin 2021.
[124] Ågren and Lindström 2014, 3.

*Concluding remarks*

TIME-US embraced both quantitative methods from labour history, and innovative digital tools to analyse historical data. This dual approach facilitated the creation of a digital corpus, enhancing the accessibility and analysis of historical documents. A key contribution of TIME-US was the datafication of historical sources. This involved collecting digitized archives, extracting and transcribing text, structuring it into XML-TEI, and publishing it online. NLP played a crucial role in this process, enabling the systematic identification and annotation of text segments related to labour activities. This way TIME-US contributed significantly to labour history by producing new quali-quantitative data and making previously inaccessible sources available for research. It highlighted the importance of digital humanities methods in historical research and demonstrated the potential of digital humanities to transform the study of labour history.

In addition to its contribution to the history of labour in France, it seems to us that TIME-US has demonstrated the full richness of interdisciplinary methods in history. From the point of view of digital humanities, the project shows the mutual benefits that historians and computer scientists can derive from ongoing collaboration: on the one hand, historians can really take advantage of advances in computer science research, and make full use of the vast amount of data available to them; on the other, computer scientists can put their techniques to the test on data from the real world, i.e. complex and often 'messy' data. TIME-US is a good illustration of the need to develop close, continuous collaboration between history and computer science. But this means that computer scientists need to be curious about historical issues and 'learn to understand' the needs of their non-specialist colleagues; historians also need to be prepared to 'engage with technology in one way or another'.[125] In our opinion, they need to go even further, by becoming actively involved in the design of the tools they use to study these large bodies of digitised text, and becoming involved in digital humanities; otherwise, they run the risk of having methods imposed on them that are incompatible with their work.[126] As Stéphane Lamassé and Philippe Rygiel rightly explain, historians need to have some control over the tools they use to carry out their research: without this, it is impossible to truly control the various stages of their work, from the collection of sources to their synthesis, description and criticism.[127]

TIME-US has shown how important it is to place historians at the centre of the process, as it is they who pose the research question, from which arise the challenges that computer scientists will help to meet. What we are talking about here is real collaboration and exchange between specialists in these two fields, and not 'service provision'. For Stéphane Lamassé and Philippe Rygiel,

> Si [...] celui-ci veut demeurer acteur d'une chaîne de production de savoir, il ne peut accepter que la constitution des corpus, la structuration et la manipulation des données, soient abandonnées à des prestataires extérieurs dont les logiques et les choix lui seraient totalement impénétrables.

> If [...] they want to remain active players in the knowledge production chain, they cannot accept that the creation of corpora and the structuring and manipulation of data should be left to external service providers whose logic and choices are totally impenetrable to them.[128]

The novelty of the TIME-US project meant that we had to invent new ways of working together. This was not achieved without trial and error and a good deal of experimentation. Let's not forget the *Annales*' conviction that the 'current historiographical moment' is a propitious time for 'an experimentation of approaches'.[129] In this respect, TIME-US is even more the heir to quantitative history and the Ecole des Annales. In *Combats pour l'histoire* (1953), Lucien Febvre called for

> Entre disciplines proches ou lointaines, négocier perpétuellement des alliances nouvelles ; sur un même sujet concentrer en faisceau la lumière de plusieurs sciences hétérogènes : tâche primordiale, et de toutes celles qui s'imposent à une histoire impatiente des frontières et des cloisonnements, la plus pressante sans doute et la plus féconde.

> the perpetual negotiation of new alliances between disciplines, whether close or distant, and the concentration of the light of several heterogeneous sciences on the same subject: this is a primordial task, and of all those that are imposed on a history impatient with frontiers and compartmentalisation, it is undoubtedly the most pressing and the most fruitful.[130]

---

[125] Salmi 2021, 55
[126] Clavert and Noiret 2013, 24-25.
[127] Lamassé and Rygiel 2014.
[128] Translated by the author. Lamassé and Rygiel 2014.
[129] Annales 2015, 216.

Whether through the borrowing of 'concepts' or 'methods and spirit', Lucien Febvre also called for collaboration, thanks to 'travailleurs d'éducation diverse s'unissant en équipes pour joindre leurs efforts' ('workers from different educational backgrounds joining together in teams to pool their efforts').[131] In 1953, he spoke of physicists, mathematicians and astronomers, and at present we can add, of computer scientists.[132] It seems to us, however, that TIME-US has perfectly embodied this 'formula for the future' as Lucien Febvre called it, by choosing to give life to a project fully rooted in the digital humanities. In doing so, it has paved the way and provided some technical tools, for other projects in France which, like it, wish to work on large digitised historical corpora[133]; we hope that it will also inspire and stimulate new research in the field of digital history.

**Bibliographical References**

---

[130] Translated by the author. Febvre 1953, 14.
[131] Translated by the author. Febvre 1953, 14.
[132] This was his opening lecture at the Collège de France on 13 December 1933. The discipline of computer science did not yet exist.
[133] For example, it directly inspired the AGODA project. See Puren, Pellet, Bourgeois, Vernus and Lebreton 2022.